\documentclass[final,5p,times,twocolumn]{elsarticle}
\usepackage{graphicx}
\usepackage{siunitx}
\usepackage{color}
\usepackage{amssymb}
\usepackage{amsmath}
\usepackage{xspace}
\usepackage{float}
\usepackage{multirow}   % Allows table elements to span several rows.
\usepackage{booktabs}   % Improves the typesettings of tables.
\usepackage{longtable}
\newcommand{\vect}[1]{\mathbf{#1}} % vector notation

\newcommand{\etal}{\textit{et al.}\xspace}
\newcommand{\ie}{\textit{i.e.,}\xspace}

\newcommand*\chem[1]{\ensuremath{\mathrm{#1}}}
\newcommand{\ku}{\ensuremath{K_\mathrm{u}}\xspace}
\newcommand{\ms}{\ensuremath{M_\mathrm{s}}\xspace}
\newcommand{\js}{\ensuremath{J_\mathrm{s}}\xspace}
\newcommand{\ax}{\ensuremath{A_\mathrm{x}}\xspace}
\newcommand{\hb}{\ensuremath{H_\mathrm{bias}}\xspace}
\newcommand{\ha}{\ensuremath{H_\mathrm{ani}}\xspace}
\newcommand{\hc}{\ensuremath{H_\mathrm{c}}\xspace}
\newcommand{\kb}{\ensuremath{k_\mathrm{B}}\xspace}
\newcommand{\mub}{\ensuremath{\mu_\mathrm{B}}\xspace}
\newcommand{\mxp}{\ensuremath{\hat{m}_x}\xspace}

\usepackage{hyperref}
\hypersetup{colorlinks=true, linkcolor=black, citecolor=black, urlcolor=black}
\usepackage{lineno}
\biboptions{comma, square, sort&compress}
\journal{npj Computational Materials}
\begin{document}
\begin{frontmatter}

\title{Full-Spin-Wave-Scaled Finite Element Stochastic Micromagnetism: Mesh-Independent FUSSS LLG Simulations of Ferromagnetic Resonance and Reversal}
\author[a]{Harald Oezelt}\corref{cor1}
\ead{harald.oezelt@gmail.com}
\cortext[cor1]{Corresponding author} 
%\cortext[cor1]{\textit{Corresponding author:} \texttt{harald.oezelt@fhstp.ac.at}}
\author[b]{Luman Qu}
\author[a,c]{Alexander Kovacs}
\author[a,c]{Johann Fischbacher}
\author[a,c]{Markus Gusenbauer}
\author[a]{Roman Beigelbeck}
%\author[d]{Bernhard Stiftner}
\author[d]{Dirk Praetorius}
\author[e]{Masao Yano}
\author[e]{Tetsuya Shoji}
\author[e]{Akira Kato}
\author[f]{Roy Chantrell}
\author[g]{Michael Winklhofer}
\author[b]{Gergely T. Zimanyi}
\author[a,c]{Thomas Schrefl}

\address[a]{Department for Integrated Sensor Systems, Danube University Krems, 2700 Wiener Neustadt, Austria}
\address[b]{Department of Physics, University of California, Davis, California 95616, USA}
\address[c]{Christian Doppler Laboratory for Magnet design through physics informed machine learning, 2700 Wiener Neustadt, Austria}
\address[d]{Institute for Analysis and Scientific Computing, Vienna University of Technology, 1040 Wien, Austria}
\address[e]{Toyota Motor Corp., Toyota City 471-8572, Japan}
\address[f]{Department of Physics, University of York, York YO10 5DD, United Kingdom}
\address[g]{Institute for Biology and Environmental Sciences IBU, Carl von Ossietzky University of Oldenburg, 26129 Oldenburg, Germany}


\begin{abstract}
In this paper, we address the problem that standard stochastic Landau-Lifshitz-Gilbert (sLLG) simulations typically produce results that show unphysical mesh-size dependence. The root cause of this problem is that the effects of spin wave fluctuations are ignored in sLLG. We propose to represent the effect of these fluctuations by a "FUll-Spinwave-Scaled Stochastic LLG", or FUSSS LLG method. In FUSSS LLG, the intrinsic parameters of the sLLG simulations are first scaled by scaling factors that integrate out the spin wave fluctuations up to the mesh size, and the sLLG simulation is then performed with these scaled parameters. 
We developed FUSSS LLG by studying the Ferromagnetic Resonance (FMR) in \chem{Nd_2Fe_{14}B} cubes. The nominal scaling greatly reduced the mesh size dependence relative to sLLG. We further discovered that adjusting one scaling exponent by less than 10\% delivered fully mesh-size-independent results for the FMR peak. We then performed three tests and validations of our FUSSS LLG with this modified scaling. 1) We studied the same FMR but with magnetostatic fields included. 2) We simulated the total magnetization of the \chem{Nd_2Fe_{14}B} cube. 3) We studied the effective, temperature- and sweeping rate-dependent coercive field of the cubes. In all three cases we found that FUSSS LLG delivered essentially mesh-size-independent results, which tracked the theoretical expectations better than unscaled sLLG. Motivated by these successful validations, we propose that FUSSS LLG provides marked, qualitative progress towards accurate, high precision modeling of micromagnetics in hard, permanent magnets.
\end{abstract}

\end{frontmatter}

\section{\label{sec:introduction}Introduction}
Finite element micromagnetic modeling has been proven to be a reliable tool to describe many magnetic phenomena at finite temperature. Usually the micromagnetic model utilizes a form of the Landau-Lifshitz-Gilbert (LLG) equation. If the computation requires varying the temperature, as for example in simulations of heat assisted magnetic recording or permanent magnets in electric motors and generators, thermal excitations and fluctuations need to be represented in the LLG method.

The most popular approach to deal with thermal excitations in micromagnetics is to use a Landau-Lifshitz-Bloch (LLB) based equation~\cite{Garanin1997, Evans2012} which combines the LLG equations for low temperatures and the Bloch equations for high temperatures. In contrast to the LLG equations, in LLB the magnetization magnitude is no longer conserved, moreover the transversal and longitudinal components have different damping parameters. The LLB gives good results for temperatures close to and higher than the Curie temperature, but it achieves this success by introducing several additional temperature dependent parameters. At high temperatures the magnetization reverses linearly by changing its length and orientation, a process which can be perfectly described using the LLB equation~\cite{ellis2015}.
Constructing these parameters involves a considerable amount of effort, though, as it requires a multi-scale simulation approach including ab initio methods and atomistic simulations, typically necessitating additional assumptions and phenomenologies~\cite{Atxitia2017,Vogler2014}.

In this paper, we focus on the stochastic Landau-Lifshitz-Gilbert equation (sLLG). For temperatures far smaller than the Curie temperature, as in the case of permanent magnet applications, the magnetization will fluctuate at the surface of its unit sphere. Switching will not occur by linear reversal. Hence, the sLLG is a good choice. However, unless we do a scaling of the parameters, the result will strongly depend on the mesh size. The here presented FUSSS version of the LLG fixes this problem.
A key advantage of sLLG for finite temperature micromagnetism is that it requires less phenomenological considerations~\cite{Brown1963,Lyberatos1993}. In the sLLG approach the thermal perturbations are represented by adding white noise to the LLG equation, turning it into a Langevin-type stochastic differential equation~\cite{Ragusa2009}. The white noise is added to the effective field of the equation in the form of a stochastic field. 

However, simulating magnetization dynamics with sLLG still requires either an accurate computation of atomistic-level spin models~\cite{Skubic2008,Evans2014}, or an approximate finite element (FE) calculation with mesh size of atomistic length scales. The intrinsic properties serving as input parameters for such finite element calculations are fixed on the single spin level at \SI{0}{K} and are usually taken from ab initio calculations. Sadly, due to the high demand of such calculations on computing resources, simulations at the atomistic scale are limited to a sample size of few nanometers.

To calculate magnetic behavior of samples on the micromagnetic scale, usually ranging from nanometers up to a few micrometers, the mesh size has to be increased. The term mesh size is used here to describe the average edge length of the elements in the mesh. Since micromagnetism is a continuum theory, its parameters represent thermal averages over the finite elements that are much larger than the material's unit cell. Simulations that use a coarser FE mesh but retain the atomistic parameters fail to include fluctuations on length scales between the interatomic spacing and the mesh size. A class of such fluctuations is the short wavelength spin waves that reduce the magnetization of the mesh-elements, a key parameter of the sLLG simulation. Therefore, the saturation magnetization \ms computed with the sLLG equation using the atomistic magnetization on a coarse mesh is too high. Such discrepancies increase with incresasing temperatures. Many works address this problem by adopting phenomenological, effective parameters to match the sLLG simulations on a finite difference or finite element mesh with experiments. Without such adaptation, the sLLG results are reliable only at low temperatures. 

Another undesirable consequence of the naive sLLG methods that do not account for the fluctuations on intra-mesh length scales is that the results of the simulations are dependent on the mesh size: larger meshes ignore more fluctuations. This is clearly an unphysical state of affairs. 

An inspired, quantitative method to account for intra-mesh fluctuations is to construct and to use renormalization group equations to integrate out intra-mesh fluctuations and represent them with scale-dependent, renormalized mesh parameters in sLLG FE simulations~\cite{Grinstein2003, Feng2001, Victora2013, Behbahani2020}. Grinstein and Koch~\cite{Grinstein2003} demonstrated the power of this method by modeling the temperature dependence of the magnetization of iron with a Heisenberg model. They used the Heisenberg renormalization group equations to generate scale-dependent, renormalized mesh parameters, which then they used in their sLLG simulations of the magnetization of large samples. Subsequently, Kirschner \etal~\cite{Kirschner2005a} performed Monte Carlo simulations on an atomistic level and calculated the average spontaneous magnetization for much coarser cells at different temperatures. These cell/mesh size dependent macro-spins were then used in sLLG simulations on length scales two orders of magnitude larger than the unit cell. Finally, some papers use simplified versions of the renormalization group and call the procedure "coarse graining", such as Behbahani \etal~\cite{Behbahani2020} for their sLLG calculation of hysteresis loops for magnetite nanorods.

However, these renormalization group approaches still have limitations. For example, renormalization group works only in the leading logarithmic approximation, which holds only if the spin waves have a gapless, purely quadratic dispersion. The need to use this idealized dispersion forced the incorporation of the anisotropy \ku and the magnetic field $\vect{h}$ only as perturbative corrections. Further, the effect of the actual crystal symmetries on the spin wave dispersion were also ignored when the momentum integrals were performed in idealized, isotropic spheres. When used for asymptotic calculations, such as near the critical temperature and at very long length scales, these approximations are justified. However, in most sLLG simulations the system is far from criticality, and the length scales of the FE mesh are intermediate, only a few nanometers. For both of these reasons, the quadratic gapless spin wave dispersion approximation is quite poor. It is especially poor for hard, permanent magnets because of their large anisotropy \ku. It gets worse still when the magnetic fields are high. And finally, reducing the analysis to the leading logarithmic approximation is justified if the renormalization group equations are integrated out to a very long correlation length, which is the case only very close to criticality. In contrast, the justification and therefore the accuracy of this leading logarithmic approximation is limited when only a limited range of wavelengths/length scales need to be integrated out, such as in the present case, when the integration proceeds from the atomistic scale only to the mesh size. For all the above reasons, for hard magnets in high fields with typical mesh sizes, the accuracy of sLLG simulations with renormalization group-corrected parameters is limited. This is consistent with the fact that the original verification of these renormalization group ideas was performed only for the soft magnet iron, in zero field, and at temperatures approaching the critical temperature $T_\mathrm{c}$~\cite{Grinstein2003}, where the approximations are the most defensible, as properly discussed in that paper.

To address the above limitations and challenges, in the present paper we report the development of an improved renormalization group, or scaling, method. 1) Instead of the exchange-only, gapless, quadratic spin wave dispersion, forced by the leading logarithmic approximation, we retain the full spin wave dispersion with the anisotropy \ku and finite magnetic field $\vect{h}$ in it. This way we avoid treating these latter terms only perturbatively. Specifically, we use the spin wave dispersion of the hard magnetic compound \chem{Nd_2Fe_{14}B} at \SI{300}{K} temperature, determined and confirmed by experiments. 2) Using these experimental dispersions also retains the proper, real spin wave dispersion that represents the crytalline symmetry of the magnetic material. With these two modifications, we integrate out the intra-mesh spin wave fluctuations from the atomistic scale to the mesh size. This integral gives rise to a scaling factor for the magnetization parameter \ms appropriate for the sLLG mesh size. From this, we determine the scaling factors for all the other parameters of the sLLG simulation for the used mesh size, including the exchange stiffness constant \ax and the effective magnetocrystalline anisotropy constant \ku. Having noticed a small discrepancy, we seek further improvements by subsequently performing FMR simulations with our micromagnetic code~\cite{Tsiantos2003} and determine updated renormalization factors for the magnetocrystalline anisotropy constant \ku at different mesh sizes. This is done by fitting \ku to shift all FMR curves to the same, shared bias field.

We will demonstrate the efficiency of our method in the context of analyzing ferromagnetic resonance (FMR). FMR measurements are a widely used technique to investigate dynamic magnetic behavior, especially to determine damping effects in magnetization dynamics~\cite{Lenz2006}. Micromagnetic FMR simulations have been performed for granular perpendicular media for magnetic recording to extract the Gilbert damping constant~\cite{Krone2011b}. In this paper, we use our "spin-wave renormalized sLLG" method to simulate the magnetic field dependence of the FMR curves and show that the simulations are in good agreement with experiments, and reassuringly, are mesh size independent.

\section{\label{sec:methods}Methods}
To calculate magnetization dynamics at finite temperatures we use the Langevin-type stochastic LLG equation. We chose to compute the magnetic state evolution of a single cube of \chem{Nd_2Fe_{14}B}. We develop and analyze the spin wave-renormalized sLLG method by simulating two test scenarios, and use a series of mesh sizes for each scenario to test the mesh size independence of the results. 

1) The equilibrium state of the cube was simulated by initially saturating the magnet in the easy axis direction and let the cube relax without any external influence. The thermal perturbation in form of the stochastic field caused increasing deviations from the easy direction. The mean value of the magnetization's $z$-component over time $\bigl\langle M_z\bigr\rangle$ was calculated for various mesh sizes to quantify the mesh dependency. 

2) The cube was again saturated in the easy-axis-direction. Then it was exposed to an increasing opposing field which eventually reversed the magnet's magnetization at the coercive field \hc. The switching simulations were performed with different mesh sizes to evaluate the possible mesh dependency of \hc.

\subsection{\label{ssec:stochLLG}Stochastic LLG for finite-temperature micromagnetics}
The sLLG method starts with constructing the usual Landau-Lifshitz-Gilbert (LLG) equations, and then adding to the effective field $\vect{h}_\mathrm{eff}$ a field representing thermal fluctuations $\vect{h}_\mathrm{th}$:
\begin{align}
\label{eq:sLLG}
\frac{\partial\vect{m}}{\partial t} &= -\gamma'\vect{m}\times\left(\vect{h}_\mathrm{eff}+\vect{h}_\mathrm{th}\right)- \frac{\gamma'\alpha}{M_\mathrm{s}^l}\vect{m}\times\left(\vect{m}\times\left(\vect{h}_\mathrm{eff}+\vect{h}_\mathrm{th}\right)\right).
\end{align}
Here $\vect{m}$ is the magnetic unit vector, the prefactor $\gamma'= \left\vert\gamma\right\vert/(1+\alpha^2)$ contains the gyromagnetic ratio $\gamma$, $\alpha$ is the Gilbert damping constant, and $\ms^l$ is the spontaneous magnetization at the minimum scale of the micromagnetic simulation, the finite element mesh size $l$, typically a few nanometers. The total field is the sum of the effective field $\vect{h}_\mathrm{eff}$ and the stochastic field $\vect{h}_\mathrm{th}$. The effective field is the sum of the exchange field $\vect{h}_\mathrm{ex} = 2\ax^l/\ms^l\Delta\vect{m}$, the anisotropy field $\vect{h}_\mathrm{ani} = 2\ku^l/\ms^l\vect{u}\left(\vect{m}\cdot\vect{u}\right)$, the demagnetizing field $\vect{h}_\mathrm{demag}$ and the applied field $\vect{h}_\mathrm{ext}$.
\begin{align}
    \vect{h}_\mathrm{eff} &= \vect{h}_\mathrm{ex}+\vect{h}_\mathrm{ani}+\vect{h}_\mathrm{demag}+\vect{h}_\mathrm{ext}
\end{align}
Here, $\vect{u}$ is the unit vector pointing in the easy axis direction.
The stochastic field $\vect{h}_\mathrm{th}$ introduces thermal fluctuations without correlation between the spatial components, in time or space. Therefore, it is a Gaussian random process which satisfies~\cite{Garcia-Palacios1998} 
\begin{align}
\Bigl\langle \vect{h}_\mathrm{th}(t)\Bigr\rangle&=0,\\
\intertext{and has the variance}
\left\langle \vect{h}_{\mathrm{th},i}(t)\,  \vect{h}_{\mathrm{th},j}(t+\Delta t)\right\rangle &= 2D\delta_{ij}\delta(\Delta t).
\end{align}
This equation relates the strength of the thermal fluctuations 
\begin{align}
D&= \frac{\alpha}{\gamma\mu_0M_\mathrm{s}^l}\frac{\kb T}{V_\mathrm{c}}
\end{align}
to the dissipation of the system depending on the FE-cell volume $V_\mathrm{c}$. Here, $\kb=\SI{1.38e-23}{J/K}$ is the Boltzmann constant, $T$ is the temperature, and $\mu_0=\SI{4\pi e-7}{H/m}$ is the magnetic vacuum permeability. In order to calculate magnetic states we interpret the stochastic equation in the Stratonovich sense and integrate \eqref{eq:sLLG} using a mid-point scheme~\cite{Werner1997} with a time step $\Delta t=\SI{1}{fs}$. 

FE-models of a cube with $\SI{40}{nm}$ edge length were prepared with various mesh sizes from \SIrange{1}{10}{nm} using Salome~\cite{Salome} and NETGEN~\cite{Schoberl1997}. A representation of a meshed model is shown in the inset of Fig.~\ref{fig:stochrelax}. The mesh is composed of uniformly sized tetrahedral elements, which is important to obtain reliable renormalization factors. Previously tested stochastic simulations proved to be very sensitive to small changes in element size of irregular meshes. 
To make progress, the sLLG simulations need the magnet's parameters, the magnetization \ms, the exchange stiffness constant \ax, the magnetocrystalline anisotropy constant \ku, and the Gilbert damping $\alpha$. Typically, sLLG approaches adopt parameters calculated by ab initio methods on the atomic scale and use them directly as the parameters of the finite elements. In simplistic terms, present sLLG approaches assume that the magnetization of the individual finite elements is fully saturated. As such, present sLLG simulations have not taken into account how the intra-mesh spin wave fluctuations reduce the magnetization from its saturated value.

\subsection{\label{ssec:SW}Spin-wave renormalization of parameters}
In order to incorporate the effects of the intra-mesh, short wavelength spin wave fluctuations up to the mesh size $l$, Grinstein and Koch proposed to integrate out the fluctuations caused by spin waves with wavenumbers higher than $\pi/l$ by renormalization group theory. Doing so yielded scale-dependent effective parameters for the sLLG simulation~\cite{Grinstein2003}. The limitations of this approach were discussed earlier, including 1) the use of an idealized, gapless spin wave dispersion; 2) the approximate, perturbative expansion in both the magnetic field $\vect{h}$ and in the anisotropy constant \ku to only leading logarithmic order; and 3) the disregard of the crystalline symmetries of the actual material, typically resulting in a non-isotropic spin wave dispersion. As also discussed, these three approximations can be justified in soft magnets in zero field, close to the critical temperature, where long wavelength critical fluctuations dominate the physics. However, typical sLLG simulations are not performed close to criticality, in zero field, and exlusively in soft magnets. For such typical simulations, the above approximations become questionable, especially for hard magnets with large \ku values.

Motivated by these limitations, in this paper we propose a similarly-inspired, but distinct method to incorporate spin wave fluctuations. In this "full-spin-wave-scaled sLLG" method, we propose to integrate out the spin wave fluctuations by using the full, preferably experimentally verified spin wave spectrum that can have a gap, is not expanded in $\vect{h}$ and in \ku, and reflects the symmetries of the crystal. It is natural to expect that our method will achieve superior accuracy of sLLG simulations for magnets with stronger crystalline anisotropy, in higher fields, having non-negligible crystalline structures, at temperatures well below criticality. 
Taken on face value, this program starts with the atomistic magnetization $\ms^a$, determined by {\it ab initio} calculation on the length scale of the unit cell $a$. The reduction of the magnetization by spin waves of wavenumber $\vect{k}$, $\Delta M(\vect{k})$, is then integrated out with wavelengths sweeping between the atomistic scale $a$ and the mesh size $l$ to yield the length-scale dependent magnetization
\begin{align}
	\ms^l = \ms^a - \int\limits_{\pi/l}^{\pi/a}\!\Delta M(\vect{k})\,\mathrm{d}\vect{k}.
\end{align}
Using the value of magnetization change caused by a spin wave leads to
\begin{align}
M_\mathrm{s}^l = M_\mathrm{s}^a - 2\mub\int\limits_{\pi/l}^{\pi/a}\!n\left(E(\vect{k}),T\right)\mathrm{d}\vect{k},
\end{align}
where the Bose-Einstein occupation factor $n\left(E(\vect{k}),T\right)$ is
\begin{align}
n\left(E(\vect{k}),T\right)=\left[\exp\left(E(\vect{k})/\left(\kb T\right)\right)-1\right]^{-1},
\end{align}
and $E(\vect{k})$ is the full spin wave spectrum that includes the anisotropy constant \ku and the magnetic field $\vect{h}$ fully, not only in leading perturbative order. $E(\vect{k})$ also reflects the discrete symmetries of the crystal, and thus can include the wavevector $\vect{k}$ in a trigonometric function instead of simply as $\vect{k}^2$. \mub is the Bohr magneton. The ratio of the renormalized magnetization to the non-renormalized magnetization will be referred to as the scaling function, or scaling factor, $s_M(l) = \ms^l/\ms^a$.

Extending the definition of the scaling function this way {\it increases} the precision with which the spin wave fluctuations are accounted for because our method is not perturbative in the magnetic field $\vect{h}$ and in the anisotropy constant \ku, and furthermore it also incorporates the discrete symmetries of the crystal. On the other hand the justification for keeping only this term becomes less compelling because strictly speaking we are not keeping only the "leading logarithmic terms" of the standard renormalization group theory. However, the farther the simulated system is from criticality, the justification to keep only the leading logarithmic terms itself becomes less compelling anyway. Therefore, for systems away from criticality the calculation of the scaling function by retaining the non-perturbative spin wave energy dispersion with the explicit crystalline symmetries becomes a net positive improvement.

The scaling of the other parameters can be constructed by using well known relationships: the exchange stiffness constant scales as $\ax^l\propto(\ms^l)^2$~\cite[p.~16]{Kronmuller2003}, and so $\ax^l/\ax^a=s_M(l)^2$. Further, the magnetocrystalline anisotropy constant scales according to Callen and Callen's power law for uniaxial anisotropy $\ku^l \propto (\ms^l)^3$~\cite{Callen1966,Durst1986}, and thus $\ku^l/\ku^a=s_M(l)^3$. 

Once all three atomistic/microscopic parameters, $\ms^a$, $\ax^a$, and $\ku^a$ have been scaled to their effective values $\ms^l$, $\ax^l$, and $\ku^l$ to incorporate the intra-mesh spin wave fluctuations up to wavelength $l$ by their scaling factors, the sLLG simulation can be performed with the mesh size $l$. This "microscale anchored full-spin-wave-scaled sLLG" should yield mesh size-independent results. 

To implement the above steps, we started by consulting the literature for the ab initio parameters of permanent magnets of interest. For hard \chem{Nd_2Fe_{14}B} magnets, Herbst summarized the results of several ab initio calculations~\cite{Herbst1991}. Quite remarkably, the calculated values showed a substantial variation, often differing by a factor of 2 or more. For this reason, it was quite difficult to establish consensus values of the ab initio calculations for hard \chem{Nd_2Fe_{14}B} magnets. 

Forced by this situation, we looked for data supported by widespread agreement. We found this among the experimentally determined macroscopic intrinsic properties of \chem{Nd_2Fe_{14}B}~\cite{Durst1986}, taken at $T=\SI{300}{K}$. The saturation magnetization is widely agreed to be $\mu_0\ms^L=\SI{1.61}{T}$, the exchange stiffness constant $\ax^L=\SI{7.7}{pJ/m^2}$, and the magnetocrystalline anisotropy constant $\ku^L=\SI{4.3}{MJ/m^3}$. 

To build on widely accepted data, we propose that the full-spin-wave-scaled sLLG method can be applied in the reverse direction as well. When the sample magnetization is measured experimentally far away from reversal processes that take place close to \hc and involve nucleation and domain wall propagation, it is reasonable to assume that the entire difference between the experimentally measured magnetization and the magnetization on the scale of the mesh size is caused by spin wave fluctuations. Therefore, it is possible to determine the effective magnetization on the length scale of the mesh size $l$ by {\it adding back} the magnetization-reduction caused by spin waves to the experimentally measured magnetization, by integrating the contribution of spin waves with wavenumbers between $\pi/L$ and $\pi/l$, where $L$ is the macroscopic system size. We call this approach the "macroscale-anchored full-spin-wave-scaled sLLG" method. We start from the experimentally measured macroscopic magnetization $\ms^L$ measured at length scale $L$, and integrate the spin wave corrections with wavelengths larger than the mesh size $l$ {\it back in}:
\begin{align}
\ms^l = \ms^L + \int\limits_{\pi/L}^{\pi/l}\!\Delta M(\vect{k})\,\mathrm{d}\vect{k}= \ms^L + 2\mub\int\limits_{0}^{\pi/l}\!n\left(E(\vect{k}),T\right) \,\mathrm{d}\vect{k}.
\end{align}
The lower limit, $\pi/L$, was taken as zero. $E(\vect{k})$ is the spin wave spectrum of an anisotropic Heisenberg ferromagnet~\cite{charap1964spin} in an external field $\vect{h}$:
\begin{align}
E(\vect{k})=2\mub\left(\frac{2\ku^L}{\ms^L}-\mu_0 \vect{h}\right)+\frac{4\mub \ax^L}{\ms^L} Z{\left(1-\frac{1}{Z}\sum_i \cos\left(\vect{a}_i \vect{k}\right)\right)},
\end{align}
where $Z$ is the coordination number of the lattice and $\vect{a}_i$ are the nearest neighbor vectors.
Experimental measurements of the spin wave spectrum are consistent with this form~\cite{Durst1986}. As outlined in the foundational parts, $E(\vect{k})$ retained the magnetic field and the crystalline anisotropy in full, instead of perturbatively expanding in them. These factors induced a gap in the spectrum, which would have been incompatible with the standard renormalization group formulation.
 
Based on the above, the scaling function for the magnetization in this macroscale-anchored full-spin-wave-scaled sLLG takes the form: 
\begin{align}
s_M\left(l\right)=\frac{\ms^l}{\ms^L} = 1+\frac{2\mub}{\ms^L}\int\limits_{0}^{\pi/l}\!n\left(E(\vect{k}),T\right) \,\mathrm{d}\vect{k}.
\end{align}
Examples of the scaling factors, $s_M(l)$, $s_A(l)$ and $s_K(l)$ as function of the mesh size $l$ are shown in Fig. \ref{fig:scaling} and discussed there.

Once the full-spin-wave-scaled magnetization $\ms^l$, exchange constant  $\ax^l$ and anisotropy $\ku^l$ have been constructed, the sLLG method with mesh size $l$ can be used to simulate experiments that involve not only spin waves but nucleation, domain wall propagation, ferro-magnetic resonance (FMR), and other, non-trivial phenomena. This full-spin-wave-scaled sLLG method should deliver mesh size independent results, and thus should introduce a major step forward in the accuracy and utility of sLLG methods for simulating complex and challenging magnetization dynamics.

\section{\label{sec:results}Results and Discussion}
\subsection{\label{ssec:FMR}Development of FUSSS LLG by simulating ferromagnetic resonance}
Often ferromagnetic resonance (FMR) experiments or simulations are used to determine the effective damping constant in materials of interest~\cite{Krone2011b, Feng2001}. In this work, we carry out FMR simulations to critically test the above developed full-spin-wave-scaled sLLG theory by checking whether it indeed delivers results independent of the finite element mesh size. In the simulations we have adopted the often-used Gilbert damping value of $\alpha=0.01$.

\begin{figure}[hbt]
	\centering
	\def\svgwidth{\columnwidth}
	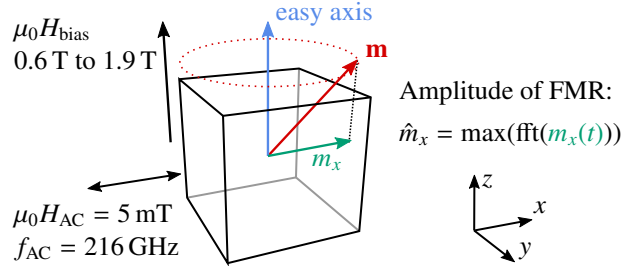
	\caption{\label{fig:fmrsetup}Scheme of setup to calculate the FMR spectra for a permanent magnetic cube with various mesh sizes. The bias field is applied in direction of the $z$ axis, \ie parallel to the magnetocrystalline anisotropy axis, and the oscillating field acts in $x$-direction. The $x$-component of the magnetization is used to calculate the response of the system for each bias field.}
\end{figure}

In the FMR-simulation, the previously prepared cube is exposed to an oscillating field with maximal amplitude of $\mu_0H_\mathrm{AC}=\SI{5}{mT}$ applied orthogonally to the cube's anisotropic easy axis in $x$-direction (see Fig.~\ref{fig:fmrsetup}). 
The frequency is chosen to be $f_\mathrm{AC}=\SI{216}{GHz}$. Neglecting the demagnetizing field, the magnetic moments are in resonance at a bias field of $\mu_0\hb=\SI{1}{T}$, resulting in an FMR-peak at this field:
\begin{align}
f_\mathrm{AC}&=\frac{\gamma\mu_0}{2\pi}\Bigl(\ha+\hb\Bigr)\label{eq:fac}\\
\text{with}\quad\ha&=\frac{2\ku^L}{\mu_0\ms^L}\label{eq:hani}\\
\text{and}\quad\gamma&=\SI{1.7608596e11}{T^{-1}s^{-1}}.
\end{align}
\ha is the theoretical anisotropy field. The bias field is applied parallel to the cube's easy axis. Initially, the cube's magnetization is saturated parallel to \hb. At a temperature of \SI{300}{K} the time evolution of the magnetization configuration is calculated by solving the stochastic LLG. The simulations are repeated with different mesh sizes and with different values of the bias field from \SIrange{0.6}{1.9}{T}. Each simulation was performed for \SI{1}{ns}. The transients settled after \SI{0.4}{ns} -- the data recorded after this transient time were accepted as the results. Sequentially the $x$-component of the magnetization, \ie the component in direction of $H_\mathrm{AC}$, was extracted from the remaining signal. The FMR curves were calculated by taking the maximal magnitude, \mxp, in the frequency spectrum of the $x$-component of $\vect{m}$.

\begin{figure}[hbt]
\centering
\def\svgwidth{\columnwidth}
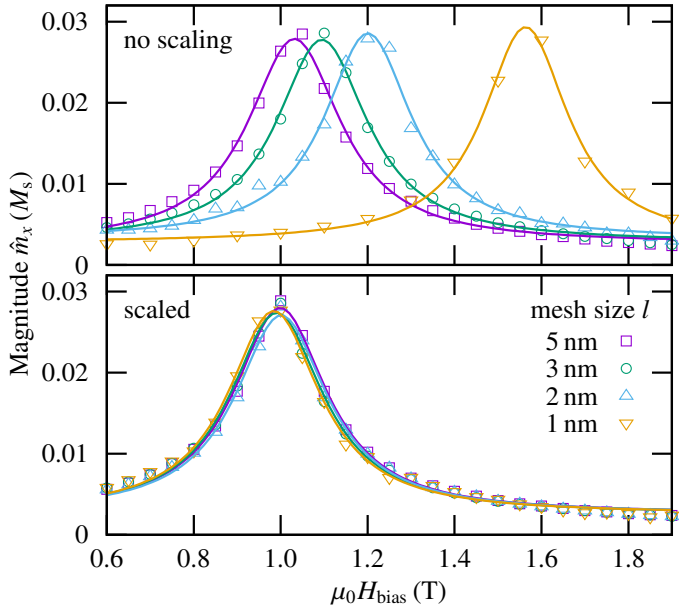
\caption{\label{fig:fmrnodemag}FMR curves obtained with sLLG without demagnetization field for different mesh sizes. The symbols mark the computed results and the solid lines are the respective fits by Lorentz functions. While without scaling (top) the curves are shifted to higher bias fields, scaling the intrinsic properties (bottom) shifts the peaks back on top of each other.}
\end{figure}

The results of various sLLG simulations for different mesh sizes are shown in Fig.~\ref{fig:fmrnodemag}. As shown in the upper panel, when the sLLG did not use parameters scaled by the spin wave fluctuations, the simulations yielded strongly mesh size dependent results. In contrast, the lower panel shows that when the full-spin-wave-scaled sLLG used parameters scaled by the spin wave fluctuations, then the simulations yielded results essentially independent of the mesh size, as discussed below.

The FMR simulations were performed for a uniformly meshed hard magnetic \chem{Nd_2Fe_{14}B} cube, wherein the mesh size $l$ was varied in the \SIrange{1}{10}{nm} range. As the upper panel shows, without scaling of the parameters, the magnetic (bias) field defining the center of the FMR peak shifted substantially from the external bias field of $\hb=\SI{1}{T}$ to the much higher bias fields of $\hb=\SI{1.6}{T}$, as $l$ varied from \SIrange{10}{1}{nm}. This large, 60\% shift demonstrates one more time the unphysical dependence of measureable quantities on the mesh size. For completeness we note that the Gilbert-damping-related FMR line-width (the FWHM of the peaks) stayed essentially invariant at $\SI{255}{mT}$~($\sigma^2=0.095$) as the mesh size $l$ was varied.

Feng and Visscher~\cite{Feng2001} introduced a mesh size dependent damping constant for a system including only exchange and thermal energy. The effective rescaled damping constant was found to increase with temperature and computational cell size. In this work we focus on permanent magnets, where in addition to the exchange the magnetocrystalline anisotropy is the dominating energy term. For the cell size of about \SI{4}{nm} and \SI{300}{K} the ratios of thermal energy to anisotropy energy and to the exchange energy are 0.015 and 0.134, respectively. Following the results of Feng and Visscher our computational experiments are in the low temperature regime where no or only minor scaling of the damping parameter is required.

When the parameters were scaled with the mesh size according to the above full-spin-wave-scaling with scaling factors $s_M(l)$, $s_A(l)=s_M(l)^2$ and $s_K(l)=s_M(l)^3$, the FMR peaks shifted back to the vicinity of the bias field of $\hb=\SI{1}{T}$. The creep of $\hb$ with mesh size was reduced from above 60\% to below 20\%, a great improvement. The fact that there was some residual dependence still left is most likely due to the fact that our scaling factors are only the leading terms of the overall spin wave fluctuation reduction of $\ms^l$. 

To eliminate even this residual mesh size dependence, we modified the scaling of $\ku$, so that the FMR peak remains at the mesh size independent $\hb=\SI{1}{T}$. To formulate the most natural scaling function for \ku, we retained the power-law form and treated its exponent as the adjustable parameter. We found that  the choice of the anisotropy scaling exponent $\ku\propto\ms^{2.72}$ kept the bias field of the FMR peak location fully mesh size independent, as shown in the bottom panel of Fig.~\ref{fig:fmrnodemag}. This is only a 9\% adjustment of the scaling exponent from its Callen value of 3, a very respectable achievement from our first order approximate scaling factors.

\begin{figure}[hbt]
\centering
\def\svgwidth{\columnwidth}
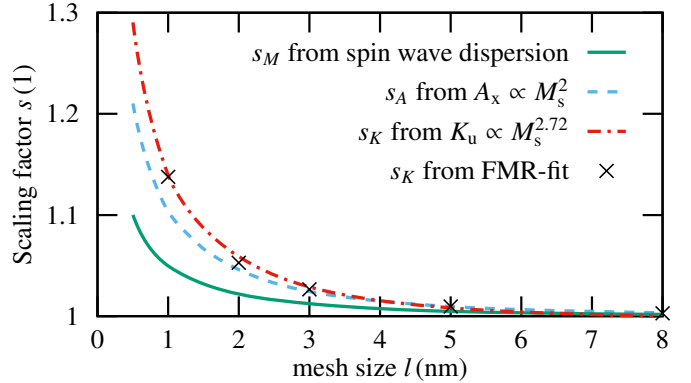
\caption{\label{fig:scaling}Scaling functions for intrinsic properties at a temperature of \SI{300}{K} for \chem{Nd_2Fe_{14}B}. While the scaling of the magnetization $s_M$ is derived from spin wave theory, the scaling for the exchange stiffness constant $s_A$ follows $\ax\propto \ms^2$. The scaling for the magnetocrystalline anisotropy constant $s_K$ was determined by choosing \ku to shift the FMR peaks of the various mesh sizes to the appropriate bias field. $s_K$ can then be fitted by $\ku^l\propto (\ms^l)^{2.72}$}
\end{figure}

The scaling functions we used for the full-spin-wave-scaled stochastic LLG are shown in Fig.~\ref{fig:scaling}. Additionally, these same scaling factors and the corresponding intrinsic property values for the used mesh sizes are listed in Table~\ref{tab:scaling}.

\begin{table}[htb]
\caption{Values of scaling functions $s$ and resulting intrinsic properties at the various mesh sizes $l$. The macroscopic intrinsic properties from experiments are shown at the bottom of the table.}
\begin{tabular*}{\columnwidth}{@{\extracolsep{\fill}}rrrrrrr}
\toprule
\toprule
\multicolumn{1}{c}{$l$} & \multicolumn{1}{c}{$s_M$} &\multicolumn{1}{c}{$s_A$} &\multicolumn{1}{c}{$s_K$} & \multicolumn{1}{c}{$\mu_0\ms^l$} & \multicolumn{1}{c}{$\ax^l$} & \multicolumn{1}{c}{$\ku^l$}\\
\multicolumn{1}{c}{\si{nm}} & \multicolumn{1}{c}{1}& \multicolumn{1}{c}{1}& \multicolumn{1}{c}{1} & \multicolumn{1}{c}{\si{T}} & \multicolumn{1}{c}{\si{pJ/m}} & \multicolumn{1}{c}{\si{MJ/m^3}}\\
\midrule
1 &1.049&1.101&1.138&1.689&8.478&4.892 \\
2 &1.023&1.046&1.053&1.647&8.054&4.527 \\
3 &1.012&1.025&1.026&1.629&7.893&4.414 \\
5 &1.005&1.010&1.010&1.618&7.777&4.341 \\
8 &1.002&1.004&1.003&1.613&7.731&4.312 \\
10&1.001&1.002&1.001&1.612&7.715&4.306 \\
\midrule
\multicolumn{4}{c}{\multirow{2}{*}{macroscopic properties}} & \multicolumn{1}{c}{$\mu_0\ms^L$} & \multicolumn{1}{c}{$\ax^L$} & \multicolumn{1}{c}{$\ku^L$}\\
& & & &\multicolumn{1}{c}{\si{T}} & \multicolumn{1}{c}{\si{pJ/m}} & \multicolumn{1}{c}{\si{MJ/m^3}}\\
\midrule
 &  & & &$1.61$ & $7.70$ & $4.30$\\
\bottomrule
\bottomrule
\end{tabular*}
\label{tab:scaling}
\end{table}

From now on, we will refer to the above developed and described FUll-Spinwave-Scaled Stochastic LLG as FUSSS LLG.

\subsection{\label{relaxation}Tests and Validation of FUSSS LLG}
\begin{figure}[hbt]
\centering
\def\svgwidth{\columnwidth}
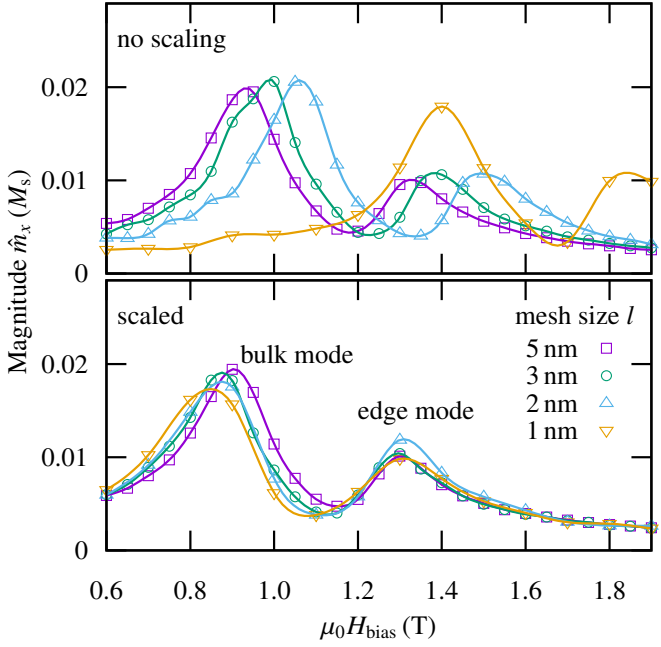
\caption{\label{fig:fmrudcub}FMR curves for different mesh sizes $l$ of a small cube computed by (top) %conventional non-thermal LLG, (center) 
sLLG without parameter scaling, and (bottom) with scaled intrinsic properties. While the symbols mark the computed results, the lines are spline interpolations serving as guide to the eye. The two peaks of each curve correspond to the bulk and edge mode.}
\end{figure}

We performed several tests and validation of our FUSSS LLG. First, we repeated the FMR-simulations by including the magnetostatic field. The FMR-curves show two peaks (see Fig.~\ref{fig:fmrudcub}). 
This curve was obtained using the fourier transform of the average magnetization of the entire cube. To analyze the spectrum we also computed the resonance of the local magnetization at different probe points. The analysis of local magnetization dynamics show that the peak at the lower \hb corresponds to the local response at the center of the cube. In this bulk mode the major part of the cube is in resonance. The second peak at higher \hb values is caused by resonance at points near the top and bottom face (near the center of the face or in the middle of an edge). The edge mode arises from a different local demagnetizing field. In addition there might be magnetostatic interaction between the modes. However, in the following we show the scaling parameters can be derived neglecting magnetostatics. 

The existence of these two modes was confirmed experimentally in FMR investigations of thin films~\cite{Zhu2010}. The two peaks were centered at $\hb \approx \SI{0.88}{T}$ and $\hb \approx \SI{1.31}{T}$, for the bulk mode and for the edge mode, respectively. As before, when the intrinsic parameters were not scaled, the sLLG curves shifted to higher bias fields as seen in the top graph of Fig.~\ref{fig:fmrudcub}. For $h=\SI{1}{nm}$ the lower field peak shifted from \SI{0.88}{T} by \SI{0.52}{T}, again a 60\% shift. To demonstrate the predictive power of our FUSSS LLG, we then repeated the simulations with the intrinsic parameters scaled by the scaling factors determined in the "no-magnetostatic-field" simulations earlier (see  Table~\ref{tab:scaling}). The bottom graph of Fig.~\ref{fig:fmrudcub} shows that the FUSSS LLG also found the two resonance peaks, and their center bias fields were almost independent of the finite element mesh size $l$.

\begin{figure}[hbt]
\centering
\def\svgwidth{\columnwidth}
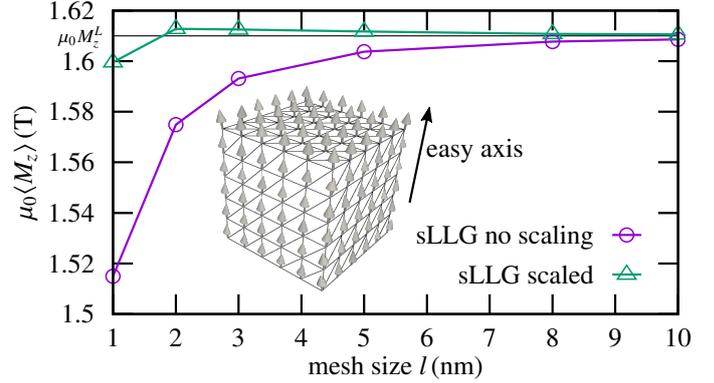
\caption{\label{fig:stochrelax}Magnetization in easy axis direction averaged over \SI{2}{ns} calculated for different mesh sizes $l$, with and without scaling of the input parameters $\ms^L$, $\ax^L$ and $\ku^L$. The inset shows an example of a mesh for the test-cube with magnetic moments in quasi-equilibrium state.}
\end{figure}

Second, we simulated the equilibrium value of the magnetization of a \chem{Nd_2Fe_{14}B} cube with an edge length of \SI{40}{nm} at \SI{300}{K} temperature. For each chosen mesh size the magnetic state was relaxed for \SI{0.5}{ns} after initial saturation in the easy axis direction. The magnetostatic field was not taken into account in these simulations. Due to the stochastic field in the sLLG, the magnetic moments fluctuated around the easy axis. In Fig.~\ref{fig:stochrelax} the mean value over \SI{2}{ns} of the magnetization component in easy direction $\mu_0\langle M_z \rangle$ is plotted against the mesh size. The macroscopic magnetization was measured to be $\mu_0M_z^L=\SI{1.61}{T}$.  Without scaling, the sLLG produces reduced values with decreasing mesh size. For $h=\SI{1}{nm}$ the mean magnetization is reduced by 6\% to \SI{1.51}{T}, as for smaller meshes the restoring force is smaller. Then we repeated the same calculation with FUSSS LLG, using scaled intrinsic parameters. As shown in the same figure, the magnetization reduction has been properly compensated, and the unphyiscal mesh-size dependence essentially eliminated. The deviations have been reduced from 6\% to less than 1\%.

\begin{figure}[hbt]
\centering
\def\svgwidth{\columnwidth}
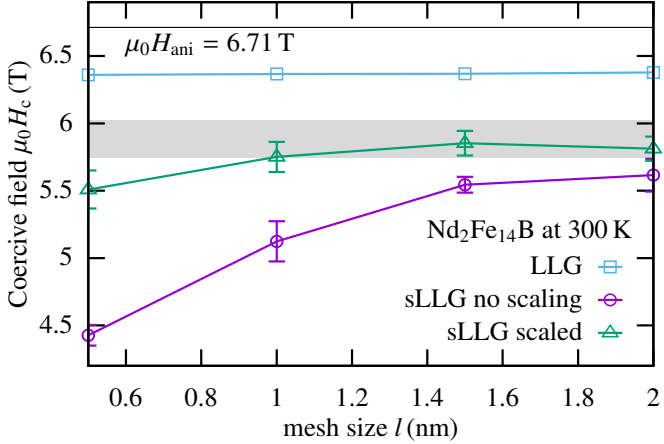
\caption{\label{fig:hc}Coercive field of a hard magnetic \SI{10}{nm} cube calculated with a field sweep rate of $v=\SI{250}{mT/ns}$. The anisotropy field is $\mu_0\ha=\SI{6.71}{T}$. Conventional, non-thermal LLG gives a mesh independent coercive field of \SI{6.36}{T}, reduced from the anisotropy field by the magnetostatic fields. Unscaled sLLG produces a coercive field heavily dependent on the mesh size, and outside the gray bend of theoretical expectations in the entire mesh-size range. In contrast, FUSSS LLG greatly reduces the mesh-size dependence, and produces a coercive field that is inside the grey band of expectations for most mesh sizes. The error bars show the standard deviation of ten independent stochastic calculations.}
\end{figure}

Third, we simulated the reversal of the magnetization by an applied field. Here we took the magnetostatic field into account. A cubic sample was magnetically saturated in the easy axis direction, and then reversed by changing the direction of the external field from the saturation direction to the opposite with a field sweep rate of $v=\SI{250}{mT/ns}$. The simulation of reversal requires a mesh size of $l\leq\sqrt{\ax/\ku}=\SI{1.34}{nm}$ to properly capture the possible formation of domain walls~\cite{Rave1998}. Therefore, only calculations for $l$ of \SIrange{0.5}{2}{nm} were performed. The edge length of the cube was shortened to \SI{10}{nm} to reduce the use of computation resources. For every mesh size the coercive field $\hc$ was extracted by averaging over 10 stochastic simulations, with and without scaling the input parameters. In Fig.~\ref{fig:hc} the mean values of \hc are shown and compared to the reversal field calculated by conventional, non-thermal LLG. 

The anisotropy field for the exchange-interaction-only model has been calculated in~\eqref{eq:hani}. With the material parameters of our cube, this comes to $\mu_0\ha=\SI{6.71}{T}$. When the magnetostatic field is incorporated, non-thermal LLG simulations show that the coercive field is reduced by \SI{0.35}{T}, to $\mu_0\hc=\SI{6.36}{T}$.

By adding thermal fluctuations, the energy barrier against switching the cube can be overcome with lower external fields. This physics can be captured as a temperature- and sweep-rate-dependent effective coercive field \hc. Various analytical expressions have been derived to describe this temperature- and sweep-rate-dependent \hc, where an applied field is swept at the constant rate $v$~\cite{Breth2012}. El-Hilo \etal~\cite{El-Hilo1992} proposed
\begin{align}
\hc&=H_0\left(1-\sqrt{ \frac{1}{\beta} \ln\left( \frac{f_0H_0}{2v\beta} \right) } \right)\\
\text{with}\quad \beta&=\frac{E_0}{\kb T}.
\end{align}
The energy barrier of the cube at zero field $E_0=\SI{522}{\kb 300K}$ was determined using the nudged elastic band method~\cite{Dittrich2002} and $H_0=\SI{6.36}{T/\mu_0}$ is the switching field without thermal fluctuations, but with magnetostatic fields. $f_0$ is the attempt frequency in zero external field, that strongly depends on the reversal mode of the sample. The approximation of $f_0$~\cite{Brown1963} for homogeneous reversal of the cube with volume $V$,
\begin{align}
f_0&=\frac{\alpha\gamma\mu_0}{1+\alpha^2}\sqrt{\frac{H_0^3\js V}{2\pi\kb T}},
\end{align}
gives an attempt frequency of $f_0=\SI{198}{GHz}$. However, recent studies of inhomogeneous reversal~\cite{Vogler2015} have suggested attempt frequencies exceeding \SI{6}{THz} for hard magnetic single particles. For these two values for $f_0$ we obtain the theoretical coercive fields of $\mu_0\hc=\SI{6.02}{T}$ and \SI{5.74}{T}, respectively. We use these two fields to define a range of expected coercive fields, signalled by the gray band in Fig.~\ref{fig:hc}. 

We first calculated the effective coercive field with the unscaled sLLG. First, the coercive field had a substantial, unphyiscal variation with changing mesh size. Second, in the entire range of mesh sizes, the coercive field \hc was well outside the grey band of theoretical expectations. Third, in fact for the smallest mesh size of $l=\SI{0.5}{nm}$, \hc was \SI{4.4}{T}, 23\% less than the lowest edge of the expected band, a substantial discrepancy.

Finally, we simulated the same reversal with FUSSS LLG, using the same scaling as in the earlier test examples. As visible in Fig.~\ref{fig:hc}, first, the mesh size dependence of \hc was largely eliminated. Second, \hc remained inside the grey band of theoretical expectations nearly the entire mesh size range. Both these facts are encouraging signs that FUSSS LLG introduces a marked, quantitative improvement over existing sLLG methods.  

\section{\label{sec:conclusion}Summary}
In this paper, we addressed the problem that standard stochastic Landau-Lifshitz-Gilbert (sLLG) simulations typically produce results that show unphysical mesh-size dependence. We identified the root cause of this problem: the effects of spin wave fluctuations are ignored in sLLG. We proposed to represent the effect of these spin wave fluctuations by a "FUll-Spin-wave-Scaled Stochastic LLG", or FUSSS LLG method. In FUSSS LLG that uses a mesh size $l$, the intrinsic parameters of the sLLG simulations \ms, \ax, and \ku are first scaled by scaling factors $s_M(l)$, $s_A(l)=s_M(l)^2$ and $s_K(l)=s_M(l)^b$ (with $b=3$) that integrate out the spin wave fluctuations up to the mesh size $l$, and the sLLG simulation is then performed with these scaled parameters. The scaling can be naturally anchored in the microscopic, or atomistic parameters. However, given that for hard magnets there is no consensus about these values, we chose to anchor our FUSSS LLG on the macroscopic scale, in experimentally measured quantities. In this sense, our scaling factors $s_M(l)$, $s_A(l)$ and $s_K(l)$ were integrating spin wave fluctuations "back in" to get the effective parameters right on the length scale of the mesh size $l$.

We developed FUSSS LLG by studying the Ferromagnetic Resonance (FMR) in \chem{Nd_2Fe_{14}B} cubes. We found that while the scaling greatly reduced the mesh size dependence relative to sLLG, full mesh size independence was not achieved with the nominal anisotropy scaling exponent of $b=3$. However, we discovered that adjusting $b$ from $b=3$ to $b=2.72$, a less than 10\% adjustment, delivered fully mesh size independent results for the FMR peak. 

We then performed three tests and validations of our FUSSS LLG with this modified scaling. First, we studied the same FMR but with magnetostatic fields included. This model exhibited two FMR peaks instead of one. Second, we simulated the total magnetization of the \chem{Nd_2Fe_{14}B} cube. Third, we studied the effective, temperature- and sweeping rate-dependent coercive field of the cubes. In all three cases we found that FUSSS LLG delivered essentially mesh-size-independent results, which tracked the theoretical expectations better than unscaled sLLG. 

Before closing, we remark that a sister version of the present FUSSS LLG can be developed for magnets where the microscopic parameters are subject of agreement by the community, in which case the microscopically anchored FUSSS LLG can be an equally powerful method.

In sum, motivated by the success of our tests and validations, we propose that FUSSS LLG provides marked, qualitative progress towards accurate, high precision modeling of micromagnetics in hard, permanent magnets.

\section*{Acknowledgement}
This work is based on results obtained from the future pioneering program "Development of magnetic material technology for high-efficiency motors" commissioned by the New Energy and Industrial Technology Development Organization (NEDO). The financial support by the Austrian Federal Ministry for Digital and Economic Affairs, the National Foundation for Research, Technology and Development and the Christian Doppler Research Association is gratefully acknowledged. The authors acknowledge the financial support by the Vienna Science and Technology fund (WWTF) under grant MA141-044. 

%\section*{References}
%\bibliographystyle{model1a-num-names}
\bibliographystyle{elsarticle-num-names}
\bibliography{references.bib}

\begin{thebibliography}{33}
\expandafter\ifx\csname natexlab\endcsname\relax\def\natexlab#1{#1}\fi
\providecommand{\url}[1]{\texttt{#1}}
\providecommand{\href}[2]{#2}
\providecommand{\path}[1]{#1}
\providecommand{\DOIprefix}{doi:}
\providecommand{\ArXivprefix}{arXiv:}
\providecommand{\URLprefix}{URL: }
\providecommand{\Pubmedprefix}{pmid:}
\providecommand{\doi}[1]{\href{http://dx.doi.org/#1}{\path{#1}}}
\providecommand{\Pubmed}[1]{\href{pmid:#1}{\path{#1}}}
\providecommand{\bibinfo}[2]{#2}
\ifx\xfnm\relax \def\xfnm[#1]{\unskip,\space#1}\fi
%Type = Article
\bibitem[{Garanin(1997)}]{Garanin1997}
\bibinfo{author}{D.~A. Garanin},
\newblock \bibinfo{title}{{Fokker-Planck and Landau-Lifshitz-Bloch equations
  for classical ferromagnets}},
\newblock \bibinfo{journal}{Physical Review B} \bibinfo{volume}{55}
  (\bibinfo{year}{1997}) \bibinfo{pages}{3050--3057}.
  \DOIprefix\doi{10.1103/PhysRevB.55.3050}.
  \href{http://arxiv.org/abs/9805054}{{\tt arXiv:9805054}}.
%Type = Article
\bibitem[{Evans et~al.(2012)Evans, Hinzke, Atxitia, Nowak, Chantrell, and
  Chubykalo-Fesenko}]{Evans2012}
\bibinfo{author}{R.~F.~L. Evans}, \bibinfo{author}{D.~Hinzke},
  \bibinfo{author}{U.~Atxitia}, \bibinfo{author}{U.~Nowak},
  \bibinfo{author}{R.~W. Chantrell}, \bibinfo{author}{O.~Chubykalo-Fesenko},
\newblock \bibinfo{title}{{Stochastic form of the Landau-Lifshitz-Bloch
  equation}},
\newblock \bibinfo{journal}{Physical Review B} \bibinfo{volume}{85}
  (\bibinfo{year}{2012}) \bibinfo{pages}{014433}.
  \DOIprefix\doi{10.1103/PhysRevB.85.014433}.
%Type = Article
\bibitem[{Ellis and Chantrell(2015)}]{ellis2015}
\bibinfo{author}{M.~Ellis}, \bibinfo{author}{R.~Chantrell},
\newblock \bibinfo{title}{Switching times of nanoscale fept: Finite size
  effects on the linear reversal mechanism},
\newblock \bibinfo{journal}{Applied Physics Letters} \bibinfo{volume}{106}
  (\bibinfo{year}{2015}) \bibinfo{pages}{162407}.
%Type = Article
\bibitem[{Atxitia et~al.(2017)Atxitia, Hinzke, and Nowak}]{Atxitia2017}
\bibinfo{author}{U.~Atxitia}, \bibinfo{author}{D.~Hinzke},
  \bibinfo{author}{U.~Nowak},
\newblock \bibinfo{title}{{Fundamentals and applications of the
  Landau–Lifshitz–Bloch equation}},
\newblock \bibinfo{journal}{Journal of Physics D: Applied Physics}
  \bibinfo{volume}{50} (\bibinfo{year}{2017}) \bibinfo{pages}{033003}.
  \DOIprefix\doi{10.1088/1361-6463/50/3/033003}.
%Type = Article
\bibitem[{Vogler et~al.(2014)Vogler, Abert, Bruckner, and Suess}]{Vogler2014}
\bibinfo{author}{C.~Vogler}, \bibinfo{author}{C.~Abert},
  \bibinfo{author}{F.~Bruckner}, \bibinfo{author}{D.~Suess},
\newblock \bibinfo{title}{{Landau-Lifshitz-Bloch equation for exchange-coupled
  grains}},
\newblock \bibinfo{journal}{Physical Review B} \bibinfo{volume}{90}
  (\bibinfo{year}{2014}) \bibinfo{pages}{214431}.
  \DOIprefix\doi{10.1103/PhysRevB.90.214431}.
  \href{http://arxiv.org/abs/1410.6066}{{\tt arXiv:1410.6066}}.
%Type = Article
\bibitem[{Brown(1963)}]{Brown1963}
\bibinfo{author}{W.~F. Brown},
\newblock \bibinfo{title}{{Thermal Fluctuations of a Single-Domain Particle}},
\newblock \bibinfo{journal}{Physical Review} \bibinfo{volume}{130}
  (\bibinfo{year}{1963}) \bibinfo{pages}{1677--1686}.
  \DOIprefix\doi{10.1103/PhysRev.130.1677}.
%Type = Article
\bibitem[{Lyberatos et~al.(1993)Lyberatos, Berkov, and
  Chantrell}]{Lyberatos1993}
\bibinfo{author}{A.~Lyberatos}, \bibinfo{author}{D.~V. Berkov},
  \bibinfo{author}{R.~W. Chantrell},
\newblock \bibinfo{title}{{A method for the numerical simulation of the thermal
  magnetization fluctuations in micromagnetics}},
\newblock \bibinfo{journal}{Journal of Physics: Condensed Matter}
  \bibinfo{volume}{5} (\bibinfo{year}{1993}) \bibinfo{pages}{8911--8920}.
  \DOIprefix\doi{10.1088/0953-8984/5/47/016}.
%Type = Article
\bibitem[{Ragusa et~al.(2009)Ragusa, D'Aquino, Serpico, Xie, Repetto, Bertotti,
  and Ansalone}]{Ragusa2009}
\bibinfo{author}{C.~Ragusa}, \bibinfo{author}{M.~D'Aquino},
  \bibinfo{author}{C.~Serpico}, \bibinfo{author}{B.~Xie},
  \bibinfo{author}{M.~Repetto}, \bibinfo{author}{G.~Bertotti},
  \bibinfo{author}{D.~Ansalone},
\newblock \bibinfo{title}{{Full micromagnetic numerical simulations of thermal
  fluctuations}},
\newblock \bibinfo{journal}{IEEE Transactions on Magnetics}
  \bibinfo{volume}{45} (\bibinfo{year}{2009}) \bibinfo{pages}{3919--3922}.
  \DOIprefix\doi{10.1109/TMAG.2009.2021856}.
%Type = Article
\bibitem[{Skubic et~al.(2008)Skubic, Hellsvik, Nordstr{\"{o}}m, and
  Eriksson}]{Skubic2008}
\bibinfo{author}{B.~Skubic}, \bibinfo{author}{J.~Hellsvik},
  \bibinfo{author}{L.~Nordstr{\"{o}}m}, \bibinfo{author}{O.~Eriksson},
\newblock \bibinfo{title}{{A method for atomistic spin dynamics simulations:
  implementation and examples}},
\newblock \bibinfo{journal}{Journal of Physics: Condensed Matter}
  \bibinfo{volume}{20} (\bibinfo{year}{2008}) \bibinfo{pages}{315203}.
  \DOIprefix\doi{10.1088/0953-8984/20/31/315203}.
  \href{http://arxiv.org/abs/0806.1582}{{\tt arXiv:0806.1582}}.
%Type = Article
\bibitem[{Evans et~al.(2014)Evans, Fan, Chureemart, Ostler, Ellis, and
  Chantrell}]{Evans2014}
\bibinfo{author}{R.~F.~L. Evans}, \bibinfo{author}{W.~J. Fan},
  \bibinfo{author}{P.~Chureemart}, \bibinfo{author}{T.~A. Ostler},
  \bibinfo{author}{M.~O.~A. Ellis}, \bibinfo{author}{R.~W. Chantrell},
\newblock \bibinfo{title}{{Atomistic spin model simulations of magnetic
  nanomaterials}},
\newblock \bibinfo{journal}{Journal of Physics: Condensed Matter}
  \bibinfo{volume}{26} (\bibinfo{year}{2014}) \bibinfo{pages}{103202}.
  \DOIprefix\doi{10.1088/0953-8984/26/10/103202}.
  \href{http://arxiv.org/abs/1310.6143}{{\tt arXiv:1310.6143}}.
%Type = Article
\bibitem[{Grinstein and Koch(2003)}]{Grinstein2003}
\bibinfo{author}{G.~Grinstein}, \bibinfo{author}{R.~H. Koch},
\newblock \bibinfo{title}{{Coarse Graining in Micromagnetics}},
\newblock \bibinfo{journal}{Physical Review Letters} \bibinfo{volume}{90}
  (\bibinfo{year}{2003}) \bibinfo{pages}{207201}.
  \DOIprefix\doi{10.1103/PhysRevLett.90.207201}.
%Type = Article
\bibitem[{Feng and Visscher(2001)}]{Feng2001}
\bibinfo{author}{X.~Feng}, \bibinfo{author}{P.~B. Visscher},
\newblock \bibinfo{title}{{Coarse-graining Landau–Lifshitz damping}},
\newblock \bibinfo{journal}{Journal of Applied Physics} \bibinfo{volume}{89}
  (\bibinfo{year}{2001}) \bibinfo{pages}{6988--6990}.
  \DOIprefix\doi{10.1063/1.1355328}.
%Type = Article
\bibitem[{Victora and Huang(2013)}]{Victora2013}
\bibinfo{author}{R.~H. Victora}, \bibinfo{author}{P.-W. Huang},
\newblock \bibinfo{title}{{Simulation of Heat-Assisted Magnetic Recording Using
  Renormalized Media Cells}},
\newblock \bibinfo{journal}{IEEE Transactions on Magnetics}
  \bibinfo{volume}{49} (\bibinfo{year}{2013}) \bibinfo{pages}{751--757}.
  \DOIprefix\doi{10.1109/TMAG.2012.2219300}.
%Type = Article
\bibitem[{Behbahani et~al.(2020)Behbahani, Plumer, and
  Saika-Voivod}]{Behbahani2020}
\bibinfo{author}{R.~Behbahani}, \bibinfo{author}{M.~L. Plumer},
  \bibinfo{author}{I.~Saika-Voivod},
\newblock \bibinfo{title}{{Coarse-graining in micromagnetic simulations of
  dynamic hysteresis loops}},
\newblock \bibinfo{journal}{Journal of Physics: Condensed Matter}
  \bibinfo{volume}{32} (\bibinfo{year}{2020}) \bibinfo{pages}{35LT01}.
  \DOIprefix\doi{10.1088/1361-648X/ab8c8d}.
%Type = Article
\bibitem[{Kirschner et~al.(2005)Kirschner, Schrefl, Dorfbauer, Hrkac, Suess,
  and Fidler}]{Kirschner2005a}
\bibinfo{author}{M.~Kirschner}, \bibinfo{author}{T.~Schrefl},
  \bibinfo{author}{F.~Dorfbauer}, \bibinfo{author}{G.~Hrkac},
  \bibinfo{author}{D.~Suess}, \bibinfo{author}{J.~Fidler},
\newblock \bibinfo{title}{{Cell size corrections for nonzero-temperature
  micromagnetics}},
\newblock \bibinfo{journal}{Journal of Applied Physics} \bibinfo{volume}{97}
  (\bibinfo{year}{2005}) \bibinfo{pages}{10E301}.
  \DOIprefix\doi{10.1063/1.1846411}.
%Type = Article
\bibitem[{Tsiantos et~al.(2003)Tsiantos, Schrefl, Scholz, and
  Fidler}]{Tsiantos2003}
\bibinfo{author}{V.~D. Tsiantos}, \bibinfo{author}{T.~Schrefl},
  \bibinfo{author}{W.~Scholz}, \bibinfo{author}{J.~Fidler},
\newblock \bibinfo{title}{{Thermal magnetization noise in submicrometer spin
  valve sensors}},
\newblock \bibinfo{journal}{Journal of Applied Physics} \bibinfo{volume}{93}
  (\bibinfo{year}{2003}) \bibinfo{pages}{8576--8578}.
  \DOIprefix\doi{10.1063/1.1557853}.
%Type = Article
\bibitem[{Lenz et~al.(2006)Lenz, Wende, Kuch, Baberschke, Nagy, and
  J{\'{a}}nossy}]{Lenz2006}
\bibinfo{author}{K.~Lenz}, \bibinfo{author}{H.~Wende},
  \bibinfo{author}{W.~Kuch}, \bibinfo{author}{K.~Baberschke},
  \bibinfo{author}{K.~Nagy}, \bibinfo{author}{A.~J{\'{a}}nossy},
\newblock \bibinfo{title}{{Two-magnon scattering and viscous Gilbert damping in
  ultrathin ferromagnets}},
\newblock \bibinfo{journal}{Physical Review B} \bibinfo{volume}{73}
  (\bibinfo{year}{2006}) \bibinfo{pages}{144424}.
  \DOIprefix\doi{10.1103/PhysRevB.73.144424}.
%Type = Article
\bibitem[{Krone et~al.(2011)Krone, Albrecht, and Schrefl}]{Krone2011b}
\bibinfo{author}{P.~Krone}, \bibinfo{author}{M.~Albrecht},
  \bibinfo{author}{T.~Schrefl},
\newblock \bibinfo{title}{{Micromagnetic simulation of ferromagnetic resonance
  of perpendicular granular media: Influence of the intergranular exchange on
  the LandauLifshitzGilbert damping constant}},
\newblock \bibinfo{journal}{Journal of Magnetism and Magnetic Materials}
  \bibinfo{volume}{323} (\bibinfo{year}{2011}) \bibinfo{pages}{432--434}.
  \DOIprefix\doi{10.1016/j.jmmm.2010.09.038}.
%Type = Article
\bibitem[{Garc{\'{i}}a-Palacios and L{\'{a}}zaro(1998)}]{Garcia-Palacios1998}
\bibinfo{author}{J.~L. Garc{\'{i}}a-Palacios}, \bibinfo{author}{F.~J.
  L{\'{a}}zaro},
\newblock \bibinfo{title}{{Langevin-dynamics study of the dynamical properties
  of small magnetic particles}},
\newblock \bibinfo{journal}{Physical Review B} \bibinfo{volume}{58}
  (\bibinfo{year}{1998}) \bibinfo{pages}{14937--14958}.
  \DOIprefix\doi{10.1103/PhysRevB.58.14937}.
%Type = Article
\bibitem[{Werner and Drummond(1997)}]{Werner1997}
\bibinfo{author}{M.~Werner}, \bibinfo{author}{P.~Drummond},
\newblock \bibinfo{title}{{Robust Algorithms for Solving Stochastic Partial
  Differential Equations}},
\newblock \bibinfo{journal}{Journal of Computational Physics}
  \bibinfo{volume}{132} (\bibinfo{year}{1997}) \bibinfo{pages}{312--326}.
  \DOIprefix\doi{10.1006/jcph.1996.5638}.
%Type = Article
\bibitem[{{Open Cascade}(2019)}]{Salome}
\bibinfo{author}{{Open Cascade}},
\newblock \bibinfo{title}{{Salom{\'{e}}: The Open Source Integration Platform
  for Numerical Simulation}},
\newblock \bibinfo{journal}{\url{www.salome-platform.org} (Accessed: June
  2019), Version 9.3}  (\bibinfo{year}{2019}).
%Type = Article
\bibitem[{Sch{\"{o}}berl(1997)}]{Schoberl1997}
\bibinfo{author}{J.~Sch{\"{o}}berl},
\newblock \bibinfo{title}{{NETGEN An advancing front 2D/3D-mesh generator based
  on abstract rules}},
\newblock \bibinfo{journal}{Computing and Visualization in Science}
  \bibinfo{volume}{1} (\bibinfo{year}{1997}) \bibinfo{pages}{41--52}.
  \DOIprefix\doi{10.1007/s007910050004}.
%Type = Book
\bibitem[{Kronm{\"{u}}ller and F{\"{a}}hnle(2003)}]{Kronmuller2003}
\bibinfo{author}{H.~Kronm{\"{u}}ller}, \bibinfo{author}{M.~F{\"{a}}hnle},
  \bibinfo{title}{{Micromagnetism and the Microstructure of Ferromagnetic
  Solids}}, \bibinfo{edition}{1} ed., \bibinfo{publisher}{Cambridge University
  Press}, \bibinfo{address}{Cambridge}, \bibinfo{year}{2003}.
%Type = Article
\bibitem[{Callen and Callen(1966)}]{Callen1966}
\bibinfo{author}{H.~Callen}, \bibinfo{author}{E.~Callen},
\newblock \bibinfo{title}{{The present status of the temperature dependence of
  magnetocrystalline anisotropy, and the power law}},
\newblock \bibinfo{journal}{Journal of Physics and Chemistry of Solids}
  \bibinfo{volume}{27} (\bibinfo{year}{1966}) \bibinfo{pages}{1271--1285}.
  \DOIprefix\doi{10.1016/0022-3697(66)90012-6}.
%Type = Article
\bibitem[{Durst and Kronm{\"{u}}ller(1986)}]{Durst1986}
\bibinfo{author}{K.-D. Durst}, \bibinfo{author}{H.~Kronm{\"{u}}ller},
\newblock \bibinfo{title}{{Determination of intrinsic magnetic material
  parameters of Nd2Fe14B from magnetic measurements of sintered Nd15Fe77B8
  magnets}},
\newblock \bibinfo{journal}{Journal of Magnetism and Magnetic Materials}
  \bibinfo{volume}{59} (\bibinfo{year}{1986}) \bibinfo{pages}{86--94}.
  \DOIprefix\doi{10.1016/0304-8853(86)90014-4}.
%Type = Article
\bibitem[{Herbst(1991)}]{Herbst1991}
\bibinfo{author}{J.~F. Herbst},
\newblock \bibinfo{title}{{\chem{R_2Fe_{14}B} materials: Intrinsic properties
  and technological aspects}},
\newblock \bibinfo{journal}{Reviews of Modern Physics} \bibinfo{volume}{63}
  (\bibinfo{year}{1991}) \bibinfo{pages}{819--898}.
  \DOIprefix\doi{10.1103/RevModPhys.63.819}.
%Type = Article
\bibitem[{Charap(1964)}]{charap1964spin}
\bibinfo{author}{S.~H. Charap},
\newblock \bibinfo{title}{{Spin-Wave Interactions in an Anisotropic
  Ferromagnet}},
\newblock \bibinfo{journal}{Physical Review} \bibinfo{volume}{136}
  (\bibinfo{year}{1964}) \bibinfo{pages}{A1131--A1136}.
  \DOIprefix\doi{10.1103/PhysRev.136.A1131}.
%Type = Article
\bibitem[{Zhu and McMichael(2010)}]{Zhu2010}
\bibinfo{author}{M.~Zhu}, \bibinfo{author}{R.~D. McMichael},
\newblock \bibinfo{title}{{Modification of edge mode dynamics by oxidation in
  Ni80Fe20 thin film edges}},
\newblock \bibinfo{journal}{Journal of Applied Physics} \bibinfo{volume}{107}
  (\bibinfo{year}{2010}) \bibinfo{pages}{103908}.
  \DOIprefix\doi{10.1063/1.3393966}.
%Type = Article
\bibitem[{Rave et~al.(1998)Rave, Ramst{\"{o}}ck, and Hubert}]{Rave1998}
\bibinfo{author}{W.~Rave}, \bibinfo{author}{K.~Ramst{\"{o}}ck},
  \bibinfo{author}{A.~Hubert},
\newblock \bibinfo{title}{{Corners and nucleation in micromagnetics}},
\newblock \bibinfo{journal}{Journal of Magnetism and Magnetic Materials}
  \bibinfo{volume}{183} (\bibinfo{year}{1998}) \bibinfo{pages}{329--333}.
  \DOIprefix\doi{10.1016/S0304-8853(97)01086-X}.
%Type = Article
\bibitem[{Breth et~al.(2012)Breth, Suess, Vogler, Bergmair, Fuger, Heer, and
  Brueckl}]{Breth2012}
\bibinfo{author}{L.~Breth}, \bibinfo{author}{D.~Suess},
  \bibinfo{author}{C.~Vogler}, \bibinfo{author}{B.~Bergmair},
  \bibinfo{author}{M.~Fuger}, \bibinfo{author}{R.~Heer},
  \bibinfo{author}{H.~Brueckl},
\newblock \bibinfo{title}{{Thermal switching field distribution of a single
  domain particle for field-dependent attempt frequency}},
\newblock \bibinfo{journal}{Journal of Applied Physics} \bibinfo{volume}{112}
  (\bibinfo{year}{2012}) \bibinfo{pages}{023903}.
  \DOIprefix\doi{10.1063/1.4737413}.
%Type = Article
\bibitem[{El-Hilo et~al.(1992)El-Hilo, de~Witte, O'Grady, and
  Chantrell}]{El-Hilo1992}
\bibinfo{author}{M.~El-Hilo}, \bibinfo{author}{A.~de~Witte},
  \bibinfo{author}{K.~O'Grady}, \bibinfo{author}{R.~Chantrell},
\newblock \bibinfo{title}{{The sweep rate dependence of coercivity in recording
  media}},
\newblock \bibinfo{journal}{Journal of Magnetism and Magnetic Materials}
  \bibinfo{volume}{117} (\bibinfo{year}{1992}) \bibinfo{pages}{L307--L310}.
  \DOIprefix\doi{10.1016/0304-8853(92)90085-3}.
%Type = Article
\bibitem[{Dittrich et~al.(2002)Dittrich, Schrefl, Suess, Scholz, Forster, and
  Fidler}]{Dittrich2002}
\bibinfo{author}{R.~Dittrich}, \bibinfo{author}{T.~Schrefl},
  \bibinfo{author}{D.~Suess}, \bibinfo{author}{W.~Scholz},
  \bibinfo{author}{H.~Forster}, \bibinfo{author}{J.~Fidler},
\newblock \bibinfo{title}{{A path method for finding energy barriers and
  minimum energy paths in complex micromagnetic systems}},
\newblock \bibinfo{journal}{Journal of Magnetism and Magnetic Materials}
  \bibinfo{volume}{250} (\bibinfo{year}{2002}) \bibinfo{pages}{12--19}.
  \DOIprefix\doi{10.1016/S0304-8853(02)00388-8}.
%Type = Article
\bibitem[{Vogler et~al.(2015)Vogler, Bruckner, Suess, and Dellago}]{Vogler2015}
\bibinfo{author}{C.~Vogler}, \bibinfo{author}{F.~Bruckner},
  \bibinfo{author}{D.~Suess}, \bibinfo{author}{C.~Dellago},
\newblock \bibinfo{title}{{Calculating thermal stability and attempt frequency
  of advanced recording structures without free parameters}},
\newblock \bibinfo{journal}{Journal of Applied Physics} \bibinfo{volume}{117}
  (\bibinfo{year}{2015}) \bibinfo{pages}{163907}.
  \DOIprefix\doi{10.1063/1.4918902}.

\end{thebibliography}
\end{document}